\documentclass[journal=jacsat,manuscript=article]{achemso}

\usepackage{hyperref}
\pdfoutput=1
\usepackage{graphicx}

\usepackage{tabularx} 
\newcolumntype{Y}{>{\centering\arraybackslash}X} 
\usepackage{colortbl} 

\usepackage{amsmath}
\usepackage{bm}
\usepackage{amssymb}
\usepackage{textcomp}
\usepackage{stmaryrd} 
\usepackage{ulem}
\usepackage{gensymb}

\usepackage{import}
\usepackage{color}
\usepackage{verbatim}

\usepackage{float}

\usepackage{sistyle} 

\newcommand{\tr}[1]{{\textrm{#1}}} 

\title{Rigid-layer Raman-active modes in $N$-layer Transition Metal Dichalcogenides: interlayer force constants and hyperspectral Raman imaging}

\author{Guillaume Froehlicher}
\affiliation{Universit\'e de Strasbourg, CNRS, Institut de Physique et Chimie des Mat\'eriaux de Strasbourg (IPCMS), UMR 7504, F-67000 Strasbourg, France}

\author{Etienne Lorchat}
\affiliation{Universit\'e de Strasbourg, CNRS, Institut de Physique et Chimie des Mat\'eriaux de Strasbourg (IPCMS), UMR 7504, F-67000 Strasbourg, France}

\author{Olivia Zill}
\affiliation{Universit\'e de Strasbourg, CNRS, Institut de Physique et Chimie des Mat\'eriaux de Strasbourg (IPCMS), UMR 7504, F-67000 Strasbourg, France}

\author{Michelangelo Romeo}
\affiliation{Universit\'e de Strasbourg, CNRS, Institut de Physique et Chimie des Mat\'eriaux de Strasbourg (IPCMS), UMR 7504, F-67000 Strasbourg, France}

\author{St\'ephane Berciaud}
\email{stephane.berciaud@ipcms.unistra.fr}
\affiliation{Universit\'e de Strasbourg, CNRS, Institut de Physique et Chimie des Mat\'eriaux de Strasbourg (IPCMS), UMR 7504, F-67000 Strasbourg, France}

\begin{document}
\maketitle
\newpage

\begin{abstract}
We report a comparative study of rigid layer Raman-active modes in $N$-layer transition metal dichalcogenides. Trigonal prismatic (2Hc, such as MoSe$_2$, MoTe$_2$, WS$_2$, WSe$_2$) and distorted octahedral  (1T', such as ReS$_2$ and ReSe$_2$) phases are considered. The Raman-active in-plane interlayer shear modes and out-of-plane interlayer breathing modes appear as well-defined features with wavenumbers in the range $0-40~\rm cm^{-1}$. These rigid layer modes are well-described by an elementary linear chain model from which the interlayer force constants are readily extracted. Remarkably, these force constants are all found to be of the same order of magnitude. Finally, we show that the prominent interlayer shear and breathing mode features allow high-precision hyperspectral Raman imaging of $N-$layer domains within a given transition metal dichalcogenide flake.
\end{abstract}


\begin{tocentry}
    \begin{center}
    \includegraphics[width=0.85\linewidth]{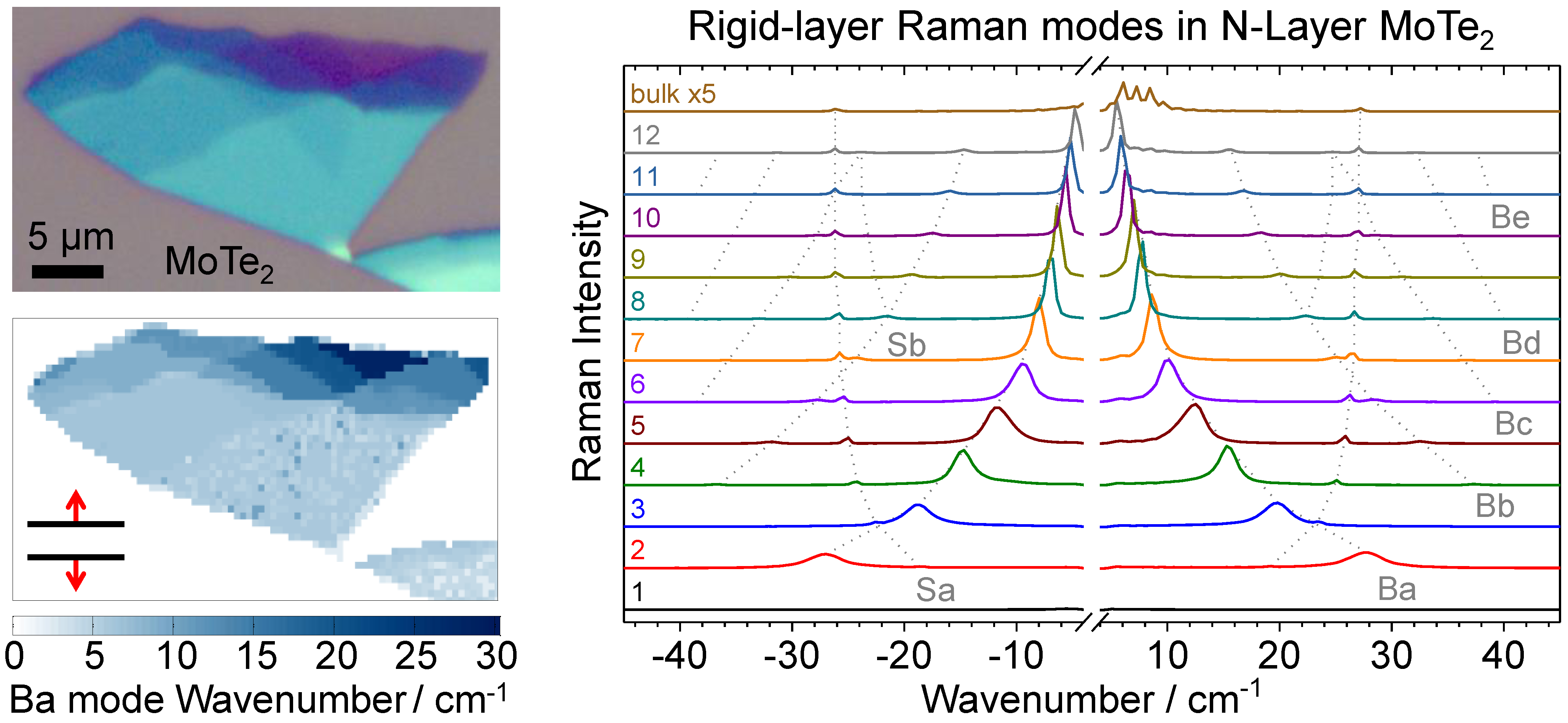}
    \end{center}
    \end{tocentry}

\newpage
Since the popularization of the mechanical exfoliation technique using adhesive tape~\cite{Novoselov2005}, the vast family of layered materials (including, graphite, boron nitride, transition metal dichalcogenides (TMDs),\dots) has attracted tremendous attention. Indeed, mechanical exfoliation and subsequent developments in direct materials growth and transfer methods allow to prepare atomically thin crystals and more recently socalled van der Waals heterostructures~\cite{Novoselov2016}. Such (quasi) two-dimensional crystals show very different properties from their three-dimensional bulk counterparts, making them ideal candidates for fundamental studies (e.g., dimensionality effects), and for various applications in nanoelectronics, optoelectronics and nanomechanics~\cite{Castroneto2009,Koppens2014,Akinwande2014,Wang2012,Xia2014,Mak2016,Castellanos2015}.

In particular, TMDs are well-documented in the bulk state~\cite{Frindt1963,Wilson1969} and currently employed for some specific applications, such as dry lubricants in the space industry~\cite{Hilton1992}. In particular, semiconducting TMDs hold great promise for the development of innovative optoelectronic devices~\cite{Koppens2014,Xia2014,Mak2016,Mueller2016}. Single layers of TMDs display exceptional, broadband optoelectronic properties that are distinct from their bulk counterparts~\cite{Mak2010,Splendiani2010}, together with a valley pseudospin that can be exploited to process information~\cite{Xu2014,Schaibley2016}. More generally, $N$-layer semiconducting TMDs offer potential advantages (including semi-transparency, low weight, large area, flexibility, strong light-matter interactions, low power consumption and scalability) over most materials currently employed in today's technologies such as silicon~\cite{Akinwande2014,Wang2012,Xia2014,Mak2016}. Research in this emerging field rely on swift, local and non-destructive characterization methods, capable of identifying the evolution of the photophysical properties with the number of layers $N$, assessing sample quality and probing interlayer interactions, including in van der Waals heterostructures. 
In this context, Raman spectroscopy has emerged as a technique of choice to characterize $N$-layer TMDs and to study in details the rich and complex exciton-phonon coupling phenomena~\cite{Lee2010,Molina2011,Luo2013,Zhang2015,Miranda2017}. In particular, as TMDs are layered compounds, they display low wavenumber ($\lesssim50~\tr{cm}^{-1}$ or equivalently $\lesssim1.5~\tr{THz}$) interlayer breathing (LBM) and shear modes (LSM). Some of these ``rigid layer modes'' have already been identified in the bulk crystals decades ago~\cite{Wieting1980}. Rigid layer modes are in first approximation  insensitive to the intralayer crystal structure, which makes them easier to model theoretically~\cite{Michel2012}. In recent years, the development of ultra-narrow notch filters has greatly facilitated low wavenumber micro-Raman measurements, in particular in two-dimensional materials~\cite{Tan2012,Plechinger2012,Zeng2012}. In TMDs, major outcomes include the experimental determination of the interlayer van der Waals force constants~\cite{Zhang2013,Zhao2013,Boukhicha2013,Froehlicher2015b}, investigation of  polytypism and in-plane anisotropy in Rhenium-based TMDs (ReS$_2$, ReSe$_2$)~\cite{Zhao2015,Nagler2015,Lorchat2016,He2016,Qiao2016}, resonant effects~\cite{Lee2015c,Soubelet2016,Kim2016,Tan2017}, as well as investigation of interlayer coupling in van der Waals heterostructures made of two distinct TMD monolayers~\cite{Lui2015}. 

In this article, to shed light on the universal behavior of rigid layer Raman modes in TMDs, and more generally in two-dimensional materials, we compare low wavenumber Raman data obtained for six different semiconducting TMDs with diverse optical bandgaps (in the range 1-2~eV) and, importantly, two distinct crystal structures: trigonal prismatic 2Hc molybdenum ditelluride (MoTe$_2$), molybdenum diselenide (MoSe$_2$), tungsten disulfide (WS$_2$) and tungsten diselenide (WSe$_2$), then distorted octohedral 1T' rhenium disulfide (ReS$_2$) and rhenium diselenide (ReSe$_2$). First, the LSM and LBM are identified as a function of the number of layer $N$ for all materials. Then, we describe the evolution of the wavenumbers of the interlayer phonon modes using a simple finite linear chain model allowing us to derive the interlayer force constants. These force constants are found to be of the same order of magnitude for all the materials. Finally, we take advantage of the sub-micrometer resolution offered by micro-Raman spectroscopy and of the existence of well-separated and intense layer rigid modes to report highly accurate hyperspectral imaging of the number of layers in TMD flakes composed of several $N$-layer-thick domains.

\section{Methods}

$N$-layer TMDs crystals were prepared by mechanical exfoliation of commercially available bulk crystals (2D semiconductors and HQ graphene) onto Si wafers covered with a 90-nm or 285-nm-thick SiO$_2$ epilayer. The number of layers was first estimated from optical contrast and atomic force microscopy measurements, and further characterized using a home-built scanning confocal micro-Raman setup. Micro-Raman scattering studies were carried out in ambient conditions, in a backscattering geometry using high numerical aperture objectives (NA=0.9 for point measurements (see Figs.~\ref{Fig1}-\ref{Fig3}) and NA= 0.6 for hyperspectral mapping (Figs.~\ref{Fig4}-\ref{Fig6}) and diffraction-limited laser spots). A monochromator equipped with a 2400 grooves/mm holographic grating and coupled to a two-dimensional liquid nitrogen cooled charge-coupled device (CCD) array was used.  Linearly polarized laser beams (at photon energies $E_{\rm L}=2.33~\rm eV$ or $E_{\rm L}=1.96~\rm eV$) were employed. Spectral resolutions of $0.4~\rm cm^{-1}$ and $0.3~\rm cm^{-1}$ were obtained at $E_{\rm L}=2.33~\rm eV$, $E_{\rm L}=1.96~\rm eV$, respectively. A laser intensity below $50~ \rm kW/cm^2$ was used in order to avoid photoinduced damage of our samples. In order to attain the low wavenumber range, a combination of one narrow bandpass filter and two ultra-narrow notch filters (Optigrate) was used. After optimization, Raman features at wavenumbers as low as 4.5~cm$^{-1}$ could be measured. Polarized Raman studies are performed using an analyzer placed before the entrance slit of our spectrometer. An achromatic half-wave plate was placed after the analyzer and adequately rotated such that the Raman backscattered beam enters the spectrometer with a fixed polarization. In the following, the sample lies in the $(x,y)$ plane and light impinges at normal incidence along the $z$ direction. Finally, the measured Raman features are fit to Voigt profiles, taking into account the spectral resolution of our apparatus. 

\section{Structural properties and phonons}

TMDs are layered crystals with chemical formula MX$_2$, where M is a transition metal atom (e.g., Mo, W, Ta, Nb, Zr,\dots) and X is a chalcogen atom (S, Se, Te)~\cite{Wilson1969}. These crystals consist in one layer of transition metal atoms sandwiched between two layers of chalcogen atoms, thus forming a X-M-X structure (see Fig.~\ref{Fig1}). Within each layer, the atoms are held together by strong covalent bonds while adjacent layers are connected by weaker van der Waals interactions. The two most common TMD polytypes are the trigonal prismatic (H) and octahedral (T) structures~\cite{Wilson1969,Katzke2004,Ribeiro2014}. These terms refer to the metal atom coordinations in the monolayer (see Fig.~\ref{Fig1}(a) and (b)). For bulk TMDs, these two polytypes are denoted 2H (since two layers are required to form the primitive unit cell) and 1T (as only one layer is required to form the bulk primitive unit cell), respectively. 

Here, we focus on 2Hc MoTe$_2$, MoSe$_2$, WS$_2$ and WSe$_2$ (i.e., hexagonal (2H) phase with /BaB AbA/ stacking~\cite{Ribeiro2014}, where upper cases represent chalcogen atoms and lower cases metal atoms) and on 1T ReS$_2$ and ReSe$_2$. More precisely, ReS$_2$ and ReSe$_2$ adopt a distorted octahedral structure (denoted 1T') with lower symmetry and significant in-plane anisotropy due to covalent bonding between Re atoms leading to quasi one-dimensional Re chains~\cite{Ho1997,Ho1998,Ho2004,Tiong1999}. Note that phase transitions between different polytypes can occur, for instance MoS$_2$ crystals can be brought in the metastable 1T/1T' phase by lithium intercalation and then brought back to the thermodynamic stable 2H phase by deintercalation due to heat~\cite{Kappera2014,Guo2015NL}.

The symmetry analysis of the phonon modes in TMDs has been thoroughly discussed in the literature, see e.g., Refs.~\citenum{Ribeiro2014,Zhang2015,Wolverson2014,Feng2015}. In brief, the point group of bulk, odd and even $N$-layer 2Hc TMDs is $D_{6h}$, $D_{3h}$ and $D_{3d}$, respectively, while it is $C_i$ for bulk, odd and even $N$-layer 1T' TMDs. From this analysis, Table~\ref{Tab1} summarizes the number of LSM and LBM, their symmetry and their activity for monolayer, bilayer, $N$-layer and bulk 2Hc and 1T' TMDs. 
 
\begin{table*} [!tbh]
\renewcommand{\arraystretch}{0.8}
\begin{center}
\small
\begin{tabularx}{0.8\textwidth}{YYYYY}
\textbf{Number of layers} &  \multicolumn{2}{c}{\textbf{2Hc TMDs}}  & \multicolumn{2}{c}{\textbf{1T' TMDs}}   \\
& \textbf{LSM} & \textbf{LBM} & \textbf{LSM} & \textbf{LBM}  \\[5pt]
\hline
\hline
\\[5pt]
1  &  $-$  & $-$ & $-$  & $-$ \\[5pt] 
\\[5pt]

2  & $\boldsymbol{E_{g}}$ & $\boldsymbol{A_{1g}}$ & 2$\boldsymbol{A_{g}}$ & $\boldsymbol{A_{g}}$ \\[5pt]

\\[5pt]

odd $N$ &  $\frac{N-1}{2}\boldsymbol{E'}$  & $\frac{N-1}{2}\boldsymbol{A'_1}$ & $(N-1)\boldsymbol{A_{g}}$ & $\frac{N-1}{2}\boldsymbol{A_{g}}$ \\[5pt]
$-$ &  $\frac{N-1}{2}E''$  & $\frac{N-1}{2}A''_2$ & $(N-1)A_u$ & $\frac{N-1}{2}A_u$    \\[5pt]

\\[5pt]

even $N$  & $\frac{N}{2}\boldsymbol{E_{ g}}$ & $\frac{N}{2}\boldsymbol{A_{ 1g}}$ & $N\boldsymbol{A_{g}}$ & $\frac{N}{2}\boldsymbol{A_{g}}$ \\[5pt]

$-$ & $\left(\frac{N}{2}-1\right)E_{u}$ & $\left(\frac{N}{2}-1\right)A_{2u}$ & $2\left(\frac{N}{2}-1\right)A_{u}$ & $\left(\frac{N}{2}-1\right)A_{u}$\\ [5pt]

\\[5pt]

bulk & $\boldsymbol{E_{2g}}$ & $B_{2g}~\star$ & $-$ & $-$ \\[5pt]
\end{tabularx}
\end{center}  
\caption{Irreducible representations of the optical phonon modes at $\Gamma$ for single-, bi-, $N$- layer and bulk 2Hc and 1T' TMDs. Bold characters denote Raman-active modes in a backscattering  geometry. Note that modes with $E''$ symmetry are Raman-active but not observable in a backscattering geometry~\cite{Loudon1964} and that modes with $E'$ symmetry are Raman- and infrared-active. Stars ($\star$) denote silent modes. All the other modes are infrared active. Note that there are no rigid layer modes in bulk 1T' TMDs because there is only one layer in the bulk primitive unit cell.}
\label{Tab1}
\end{table*}


\section{Identification of the LBM and LSM}

\begin{figure*}[!tbh]
\begin{center}
\includegraphics[width=0.8\linewidth]{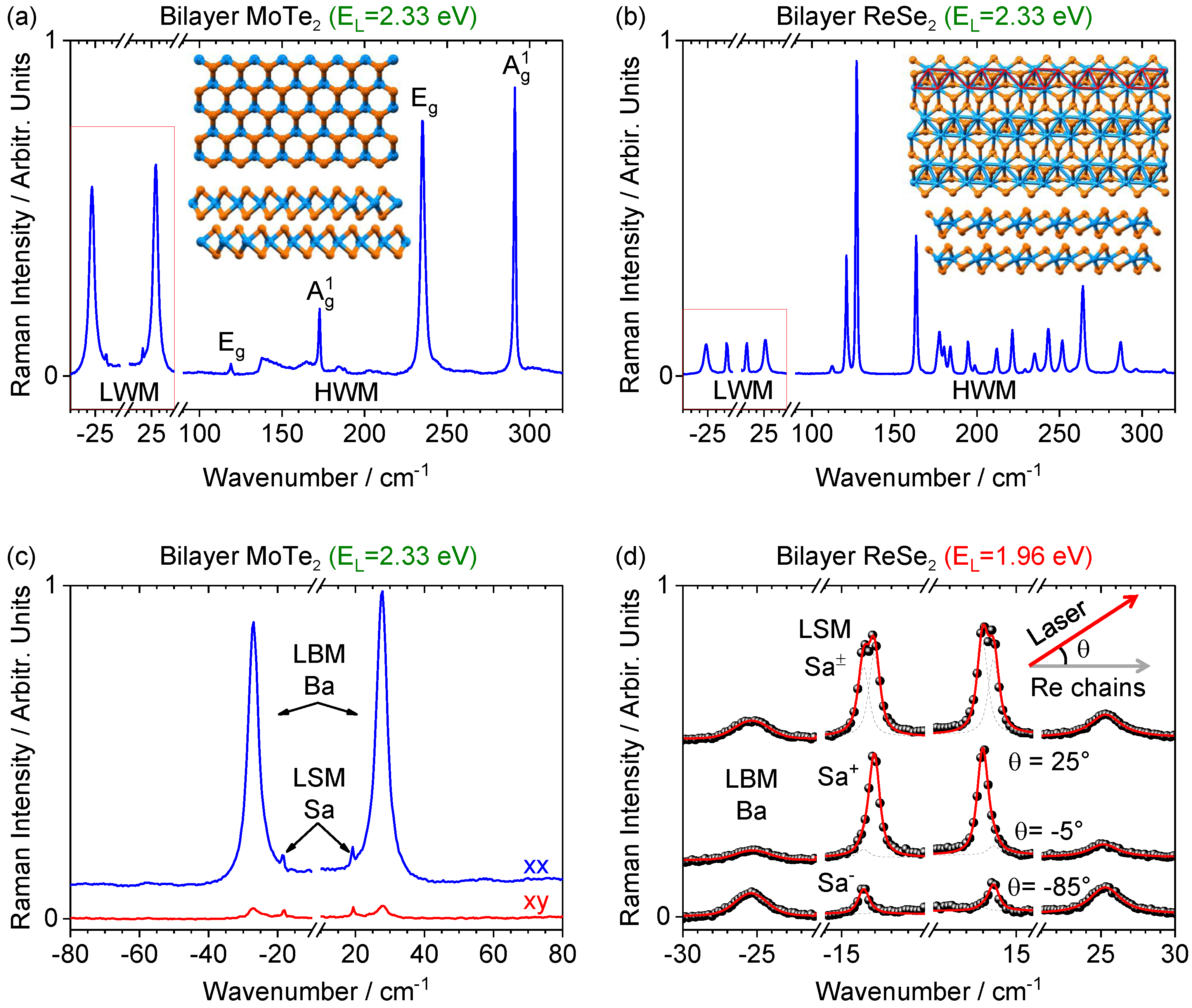}
\caption{Micro-Raman spectra of (a) bilayer MoTe$_2$ and (b) bilayer ReSe$_2$ in the parallel (left panel, $xx$) configuration. LWM and HWM denote the low wavenumber (rigid layer) modes and high-wavenumber modes, respectively. The top and side view of both structures are represented (metal and chalcogen atoms are in dark yellow and blue, respectively).   (c) Close up on the low wavenumber rigid layer modes (red boxes in (a) and (b)) in bilayer MoTe$_2$ in the parallel $(xx)$ and perpendicular $(xy)$ configurations. The limited extinction of the breathing mode (LBM) feature in the $xy$ configuration is due to slight depolarization effects in our micro-Raman setup. (d) Low wavenumber Raman spectra of bilayer ReSe$_2$ recorded in the $xx$ configuration at three distinct relative orientations of our sample with respect to incoming laser polarization. Measurements at $E_{\rm L}=1.96~\rm eV$ provide a slightly better spectral resolution and allow the observation of the LSM splitting.}
\label{Fig1}
\end{center}
\end{figure*}

\begin{figure*}[!tbh]
\begin{center}
\includegraphics[width=0.75\linewidth]{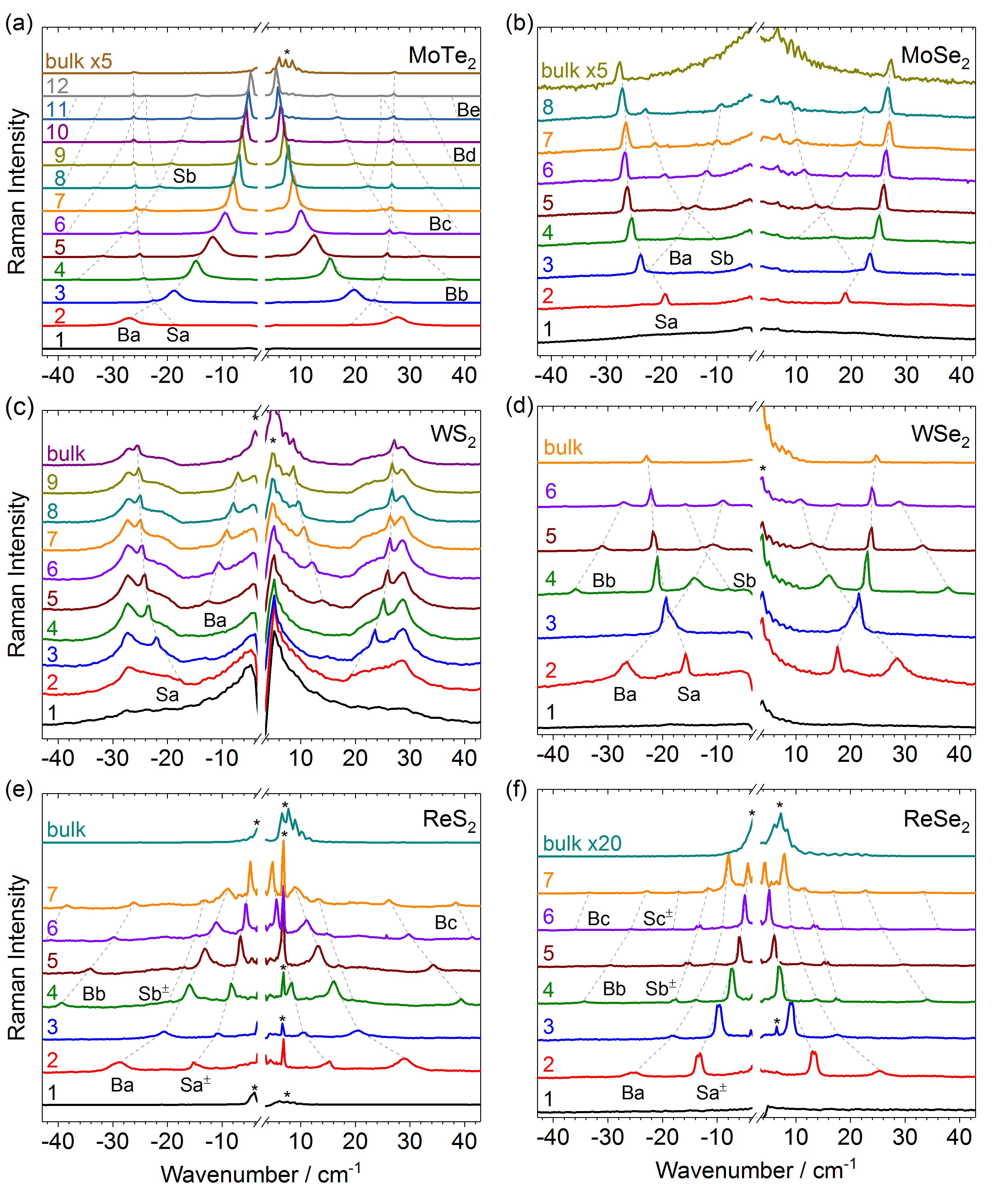}
\caption{Low wavenumber micro-Raman spectra of (a) $N=1$ to $N=12$ layer and bulk MoTe$_2$ (b) $N=1$ to $N=8$ layer and bulk MoSe$_2$,(c) $N=1$ to $N=8$ layer and bulk WS$_2$, (d) $N=1$ to $N=6$ layer and bulk WSe$_2$, (e) $N=1$ to $N=7$ layer and bulk ReS$_2$, (f) $N=1$ to $N=7$ layer and bulk ReSe$_2$. The dashed lines follow the extracted wavenumbers of the interlayer shear and breathing  modes, labelled according to the text (see also Fig.~\ref{Fig3}). The asterisks on the bulk spectra highlight residual stray light from the exciting laser beam. All spectra were recorded at $E_{L}=2.33~\rm eV$.}
\label{Fig2}
\end{center}
\end{figure*}

Figure~\ref{Fig1} shows typical Raman spectra for bilayer MoTe$_2$ (as a 2Hc compound, see Fig.~\ref{Fig1}a) and for bilayer ReSe$_2$ (as a 1T' compound, see Fig.~\ref{Fig1}b). In the following, we will exclusively discuss the low wavenumber region of the spectrum. In the case of 2Hc TMD bilayers, the LBM and LSM are readily identified by comparing the Raman spectra recorded in the parallel $(xx)$ and perpendicular $(xy)$ configurations. The LBM (with $A_{1g}$ symmetry) is drastically attenuated in the $xy$ configuration, whereas the doubly degenerate LSM (with $E_{2g}$ symmetry) has equal intensity in the $xx$ and $xy$ configurations (see Fig.~\ref{Fig1}c). In the case of 1T' TMD bilayers, both the LSM and LBM have $A_{g}$ symmetry and display non-trivial dependence on the polarization conditions. By carefully rotating the sample relative to the fixed laser polarization and recording the Raman spectra in the $xx$ configuration, it is possible to spectrally resolve the  splitting between the two LSM that arises due to in-plane anisotropy (see Fig.~\ref{Fig1}d and Ref.~\citenum{Lorchat2016}).

Figure~\ref{Fig2} display low wavenumber Raman spectra as a function of $N$ for six different TMDs. All spectra in this figure were recorded at $E_{L}=2.33~\tr{eV}$. As in previous reports~\cite{Zhang2013,Zhao2013,Boukhicha2013,Froehlicher2015b}, for $N\geq2$, one can clearly identify the previously reported branches of LBM and LSM modes, whose wavenumber vary significantly with $N$.  The LSM and LBM branches are identified using polarized measurements as described above and are denoted Ba, Bb,\dots and Sa, Sb,\dots (Sa$^\pm$, Sb$^\pm$ for ReS$_2$ and ReSe$_2$, where the subscript $\pm$ denotes the two split LSM features), respectively (see Fig.~\ref{Fig3}).
As further discussed below, the absolute and relative intensities of the LSM and LBM vary largely depending on the material and on $E_{L}$ (not shown), suggesting pronounced resonance Raman effects. Remarkably, in the case of WS$_2$, a prominent mode appears near $27~\rm{cm}^{-1}$, irrespective of $N$. This feature, also previously seen in MoS$_2$ layers~\cite{Lee2015} has tentatively been assigned to a higher order resonant Raman process involving finite momentum transverse acoustic (TA) phonons and electron-defect scattering, such that energy and momentum are conserved~\cite{Tan2017}.  

\section{Finite linear chain model}

\begin{figure*}[!th]
\begin{center}
\includegraphics[width=0.75\linewidth]{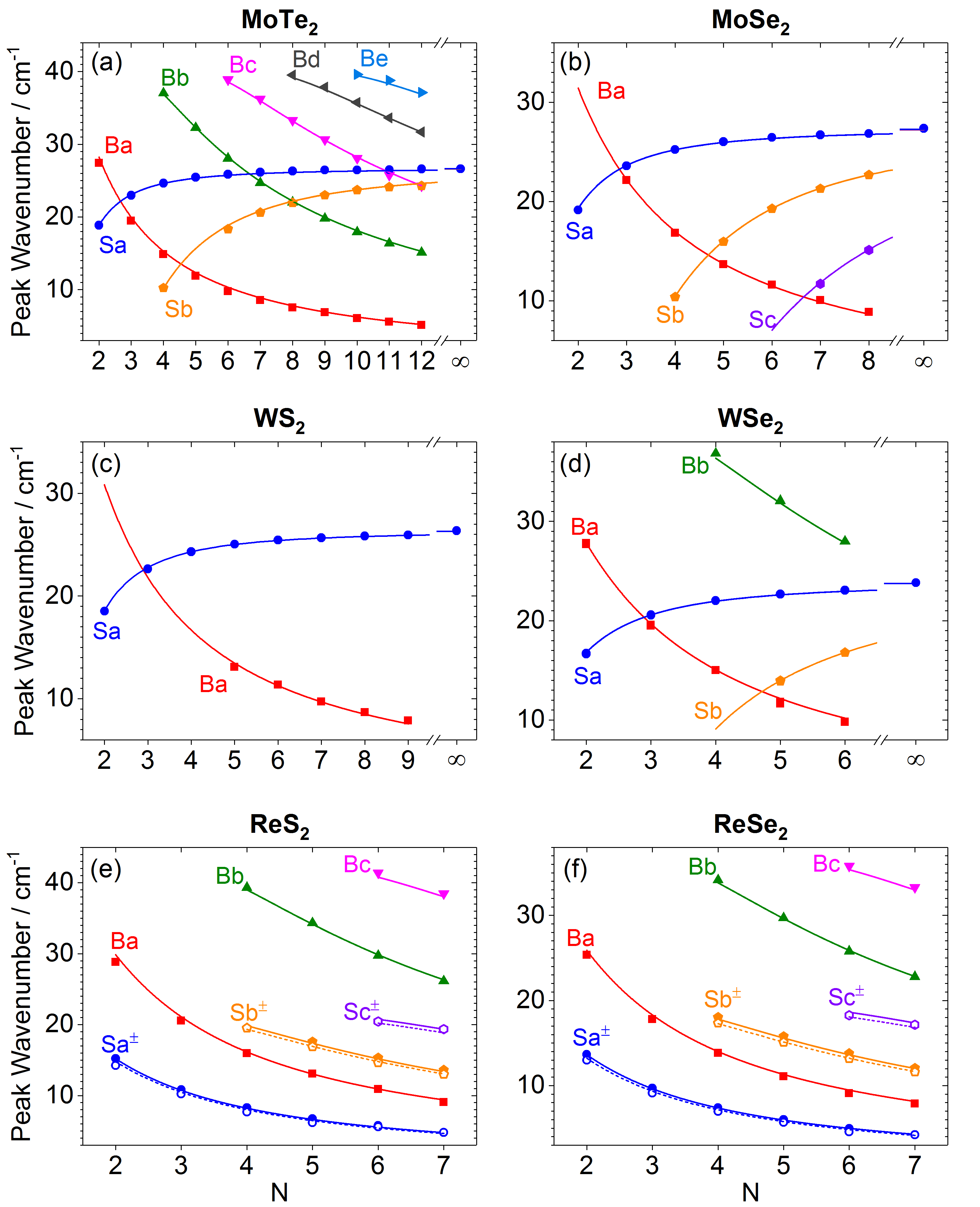}
\caption{Wavenumbers (symbols) of the LSM (Sa, Sb,\dots,) and LSM (Ba, Bb,\dots) as a function of $N$ for $N$-layer (a) MoTe$_2$, (b) MoSe$_2$, (c) WS$_2$, (d) WSe$_2$, (e) ReS$_2$ and (f) ReSe$_2$.  The measured wavenumbers are globally fit to Eq.~\eqref{eq_freq_1} (lines). The theoretical curves are plotted in the range of $N$ where the corresponding mode is observable. Due to in-plane anisotropy, each LSM branches of ReS$_2$ and ReSe$_2$ is split into a lowest- (open symbols and dashed lines) and a highest-wavenumber branch (filled symbols and solid lines), leading to two shear ($\kappa_\tr{S}$) force constants~\cite{Lorchat2016}. The extracted force constants are reported in Table~\ref{Tab2}.}
\label{Fig3}
\end{center}
\end{figure*}

\begin{figure}[!th]
\begin{center}
\includegraphics[width=0.5\linewidth]{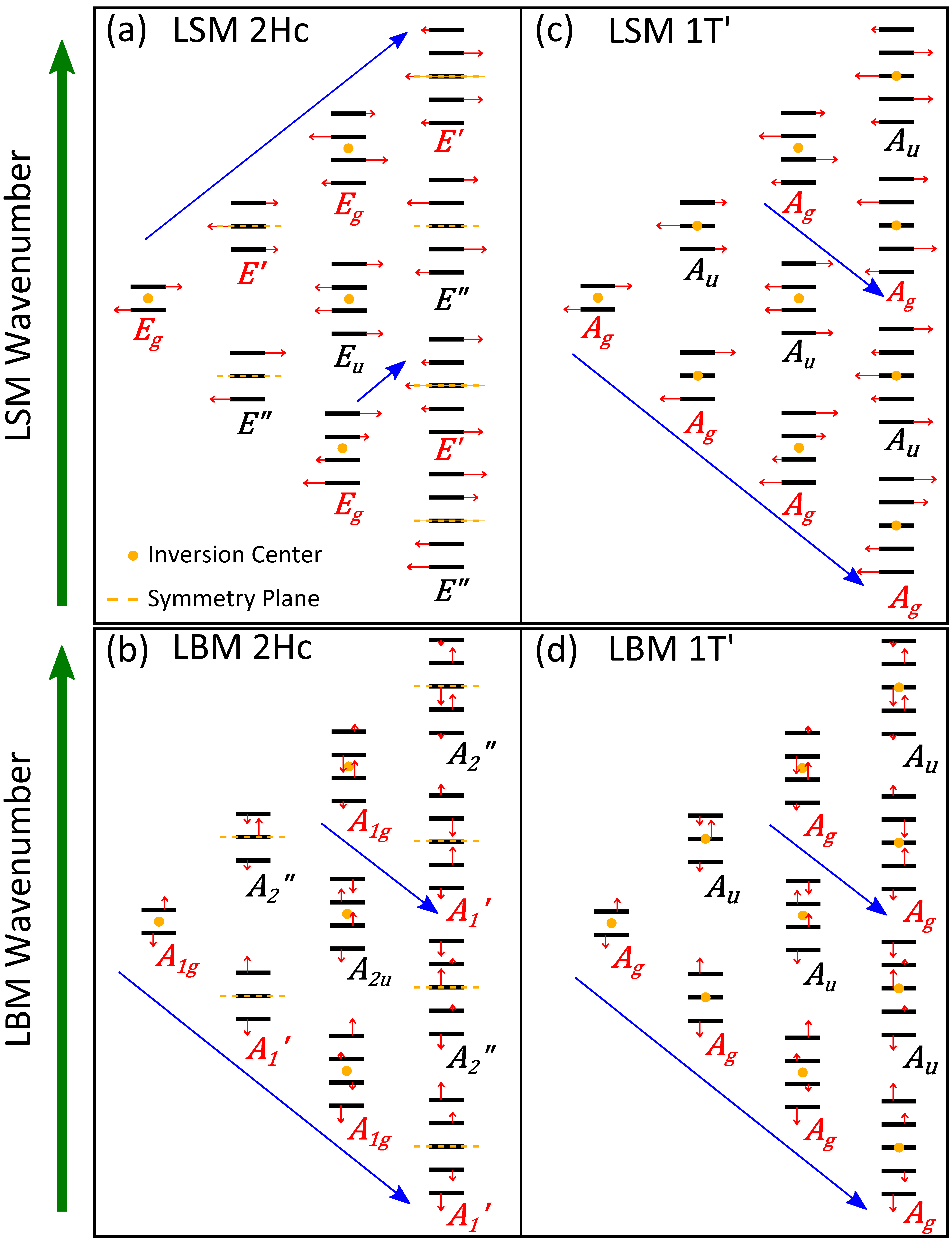}
\caption{Normal displacements corresponding to the LSM and LBM calculated using Eq.\eqref{eq_ND_2} for TMD with a 2Hc (a,b) or 1T' (c,d) structure. The presence of an inversion center (for even $N$ in the 2Hc phase and for any value of $N$ in the 1T' phase) or of a mirror symmetry plane (for odd $N$  in the 2Hc phase) is indicated. The irreducible representations are indicated, with Raman-active modes in red, infrared-active modes and Raman modes that are not observable in a backscattering geometry in black. The oblique arrows symbolize the decrease (increase) of the wavenumber of one given mode as $N$ increases. This figure is inspired by Ref.~\citenum{Lorchat2016}.}
\label{Fig3b}
\end{center}
\end{figure}

The wavenumbers of the LSM and LBM are displayed in Fig.~\ref{Fig3}. The evolution of these wavenumbers with $N$ can be derived using density functional theory (DFT)~\cite{Zhao2013,Boukhicha2013} or analytically, by diagonalizing the dynamical matrix~\cite{Michel2012,Zhang2013}. In the case of interlayer modes, we can use a finite linear chain model~\cite{Luo1996}, in which each layer is treated as a rigid mass unit with a mass $\mu=2\mu_\tr{X}+\mu_\tr{M}$, where $\mu_\tr{X}$ is the mass per unit area of the chalcogen atom and $\mu_\tr{M}$ of the metal atom, to derive the dynamical matrix. Applying this model to a $N$-layer sample and diagonalizing the dynamical matrix leads to the $N$ normal modes of this system, where the displacement of the i$^\tr{th}$ layer plane ($i\in \llbracket 1,N \rrbracket$) is $u^p_i=\varepsilon^p_i\tr{e}^{-\tr{i}\omega_p t}$ with the eigenwavenumbers $\omega_p$ and eigenvectors $\varepsilon_i^p$:
\begin{eqnarray} 
\omega_p &=& \frac{1}{2 \pi c}\sqrt{\frac{2\kappa}{\mu}\left[1-\cos\left(\frac{(p-1)\pi}{N}\right)\right]} \tr{~with~} p\in\llbracket1,N\rrbracket, \label{eq_freq_1} \\
\varepsilon^p_i &=&
\left\lbrace
\begin{array}{ccl}
\frac{1}{\sqrt{N}} & \tr{if} & p=1,\\
\sqrt{\frac{2}{N}}\cos\left(\frac{(p-1)(2i-1)\pi}{2N}\right) & \tr{if} & p\in\llbracket2,N\rrbracket,
\end{array}\right.,
\label{eq_ND_2}
\end{eqnarray}
where $\kappa$ is the interlayer force constant and $c$ denotes the speed of light. 
Note that $p=1$ corresponds to the acoustic mode with a zero wavenumber and a center of mass not at rest. For $N=2$, the non-zero wavenumber is $\omega=\frac{1}{2 \pi c}\sqrt{\frac{2\kappa}{\mu}}$ and for the bulk wavenumber $\omega=\frac{1}{2 \pi c}\sqrt{\frac{4\kappa}{\mu}}$. These two expressions can be deduced from simple considerations. For $N=2$, the problem is equivalent to a spring-mass system with a reduced mass $\mu/2$ and a spring $\kappa$. Such an oscillator has an eigenwavenumber $\omega=\frac{1}{2 \pi c}\sqrt{\frac{2\kappa}{\mu}}$. For the bulk, the two masses are connected by two springs of stiffness $\kappa$ because of the periodic boundary conditions. This system is equivalent to one oscillator with a reduced mass $\mu/2$ and a spring constant $2\kappa$ which wavenumber is $\omega=\frac{1}{2 \pi c}\sqrt{\frac{4\kappa}{\mu}}$. 

Using Eq.~\eqref{eq_freq_1}, we globally fit the LSM and LBM branches with the shear ($\kappa_\tr{S}$) and breathing ($\kappa_\tr{B}$) force constants as the only fitting parameters. The LSM and LBM normal displacements are represented in Fig.~\ref{Fig3b} up to $N=5$. Remarkably, for 2Hc TMDs, the experimentally observed Raman-active LSM Sa, Sb, Sc, \dots (resp. LBM Ba, Bb, Bc, \dots) correspond to branches $p=N,~N-2,~N-4, \dots$ (resp. $p=2,~4,~6, \dots$) in Eq.~\eqref{eq_freq_1}, with increasing (resp. decreasing) wavenumber as $N$ augments. In contrast, for 1T' TMDs, both LSM Sa$^{\pm}$, Sb$^{\pm}$, Sc$^{\pm}$, \dots~and LBM Ba, Bb, Bc, \dots wavenumbers decrease as $N$ increases and correspond to branches $p=2,~4,~6, \dots$ in Eq.~\eqref{eq_freq_1}. These contrasting behaviors have previously been rationalized using a symmetry analysis (see Fig.~\ref{Fig3b} and Ref.~\citenum{Lorchat2016}). More generally, there is no straightforward connection between the Raman activity (determined by the crystal symmetries) and the measured Raman intensity of a given Raman-active mode. For example, it has been shown that the most intense experimentally observed LBM in bernal-stacked and twisted $N$-layer graphene correspond to $p=2$(Ref.~\citenum{Lui2014}) and $p=N$ (Ref.~\citenum{Wu2015b})  in Eq.~\eqref{eq_freq_1}, respectively.

Eq.~\eqref{eq_freq_1} provides an excellent fit to our data for all materials (see Fig.~\ref{Fig3}). This result demonstrates that higher-order nearest neighbor interactions~\cite{Luo2013,Froehlicher2015b} can be neglected when describing rigid layer modes. The extracted force constants are reported in Table~\ref{Tab2} and compared to measurements on other layered materials found in the literature. Despite different compounds and crystal structure, we notice that the interlayer force constant are all very close. This conclusion is not surprising for 2Hc compounds. However, since the evolution of the optical properties of ReS$_2$ and ReSe$_2$ with varying $N$ are not as pronounced as in the case of MoX$_2$ and WX$_2$, it has been suggested that the layers that compose a ReX$_2$ stack might be weakly coupled~\cite{Tongay2014}. Low wavenumber Raman measurements rather demonstrate that ReX$_2$ should be considered as van der Waals materials with similar interlayer coupling as 2Hc TMDs~\cite{Zhao2015,Nagler2015,Lorchat2016,He2016,Qiao2016}. Let us also note that polytypism (as observed in MoS$_2$~\cite{Lee2016} and ReS$_2$~\cite{He2016,Qiao2016}) imply that stacking order with different symmetries may lead to slightly different force constants and fan diagrams for the rigid layer modes.

\begin{table*} [!tbh]
\renewcommand{\arraystretch}{1.55}
\begin{center}
\begin{tabularx}{0.9\textwidth}{YYY}
\textbf{Material} & ${\rm \bold{\boldsymbol{\kappa}_{S}~(10^{18}~N/m^3)}}$ & ${\rm \bold{\boldsymbol{\kappa}_{B}~(10^{18}~N/m^3)}}$  \\
\hline
\hline
MoTe$_2$ (Ref.~\citenum{Froehlicher2015b} and this work) & 34.2 & 76.9  \\
MoTe$_2$~ (Ref.~\citenum{Grzeszczyk2016,Song2016}) & 36.0,~42.5 & 75.0,~91.2 \\
MoSe$_2$ (this work) & 29.6 & 78.4\\
MoSe$_2$~(Ref.~ \citenum{Kim2016}) & 29.2 & 87.3 \\
WS$_2$ (this work) & 29.4 & 80.6 \\
WSe$_2$ (this work) & 30.5 & 83.7  \\
WSe$_2$~\citenum{Zhao2013} & 30.7 & 86.3   \\
ReS$_2$ (Ref.~\citenum{Lorchat2016} and this work) & 17.1/18.9 & 69.3 \\
ReSe$_2$ (Ref.~\citenum{Lorchat2016} and this work) & 17.8/19.4 & 69.2 \\
MoS$_2$~(Ref.~\citenum{Boukhicha2013,Zhang2013,Zhao2013}) & 28.1,~28.2,~27.2 & 88.1,~89.0,~86.2  \\
Graphite~(Ref.~\citenum{Tan2012,Lui2014}) & 12.8 & 88.0 \\
\end{tabularx}
\end{center}  
\caption{Force constants per unit area  extracted from the fit of the experimental data to the finite linear chain model (see Fig.~\ref{Fig3}) and from previous literature. For ReS$_2$ and ReSe$_2$, the shear force constants $\kappa_\tr{S}$ correspond to the lowest-/highest-wavenumber branch, respectively.}
\label{Tab2}
\end{table*}

\begin{figure*}[!t]
\begin{center}
\includegraphics[width=0.8\linewidth]{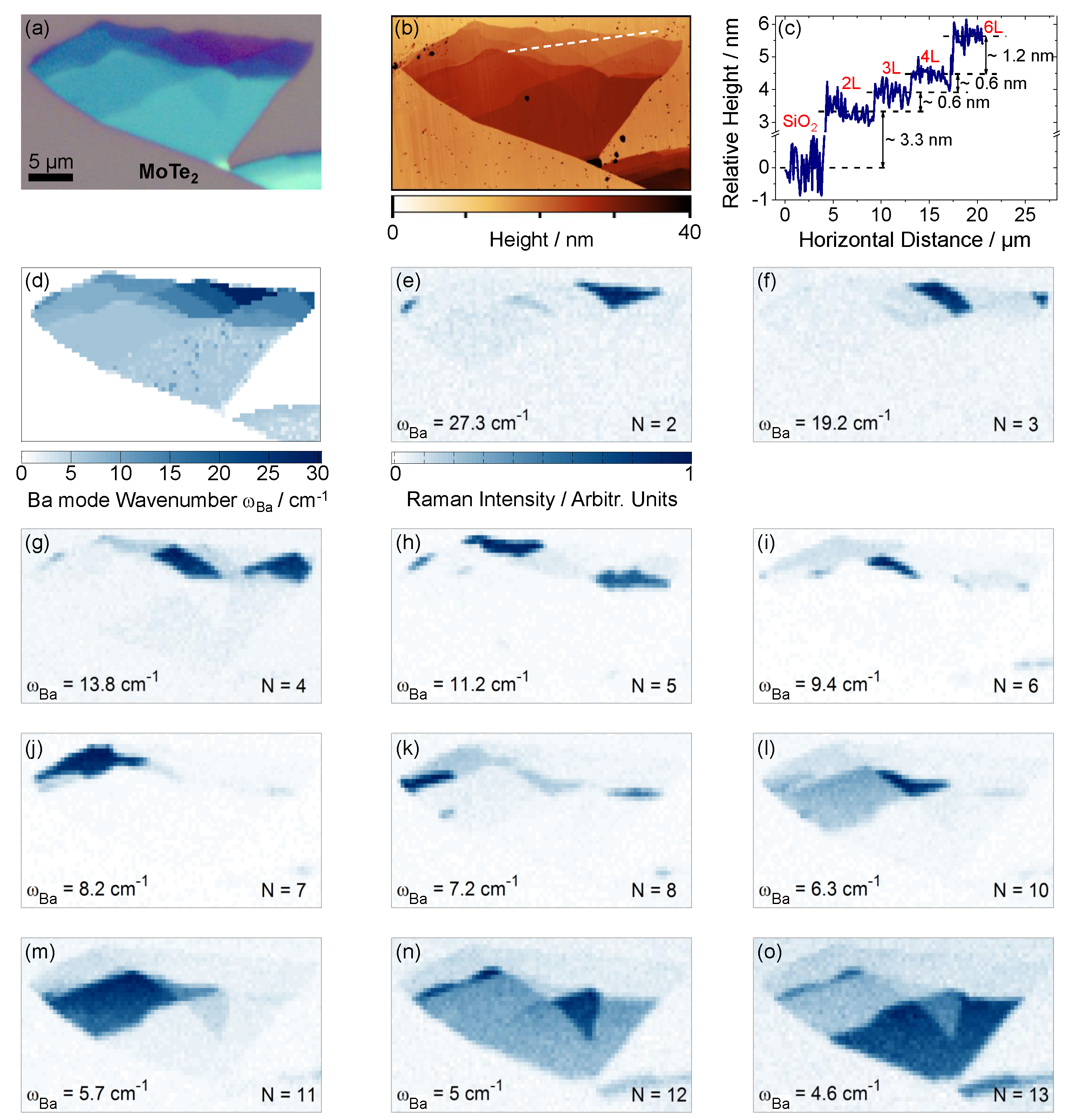}
\caption{Optical (a) and atomic force microscopy (AFM) (b) images of a MoTe$_2$ flake. A height profile along the white dashed-line shown in the AFM image is shown in (c). (d) Hyperspectral map of the wavenumber of the most prominent interlayer breathing mode feature (denoted Ba in Figs.~\ref{Fig2} and \ref{Fig3}). (e-o) Hyperspectral maps of the integrated Raman intensity at selected wavenumbers that correspond to the wavenumber of the Ba feature at a given value of $N$. Measurements are performed at a laser photon energy of 2.33 eV.}
\label{Fig4}
\end{center}
\end{figure*}

To close this section, we provide an alternative description of rigid-layer phonon modes in $N$-layer crystals using the bulk phonon dispersion relation~\cite{Karssemeijer2011,Michel2012}. This description can be viewed as a ``top-down'' approach, as opposed to a ``bottom-up'' description outlined above. The dispersion relation of an infinite monoatomic linear chain of lattice parameter $\delta$ is given by~\cite{Ashcroft1976}
\begin{equation}
\omega(q)=\frac{1}{2 \pi c}\sqrt{\frac{2\kappa}{\mu}(1-\cos q \delta)}, \label{eq_disp_mono_chain}
\end{equation}
where $q\in \left[-\frac{\pi}{\delta},\frac{\pi}{\delta}\right]$ is the phonon wavevector taken in the Brillouin zone of the monoatomic linear chain. 
However, Eq.~\eqref{eq_disp_mono_chain} is not the dispersion relation of the bulk crystal since the unit cell contains two layers. Nevertheless, it can be deduced from Eq.~\eqref{eq_disp_mono_chain} knowing that the size of the Brillouin zone changes from $\frac{2\pi}{\delta}$ to $\frac{\pi}{\delta}$. Hence the dispersion relation of the bulk crystal is
\begin{equation}
\omega(q)=\frac{1}{2 \pi c}\sqrt{\frac{2\kappa}{\mu}(1\pm \cos q\delta)}, \label{eq_disp_bulk}
\end{equation}
with $q\in\left[-\frac{\pi}{2\delta},\frac{\pi}{2\delta}\right]$. The phonon branch with the $+$ ($-$) corresponds to the optical (acoustic) branch.  Note that $q=0$ gives the two wavenumbers of the bulk zone center phonon modes. For non-zero wavevector $q$ and $N\geq2$~\footnote{For $N=1$, $q=0$ and thus the only mode is the acoustic one.}, comparing Eqs.~\eqref{eq_freq_1} and \eqref{eq_disp_bulk} yield
\begin{equation}
q(\nu)=\frac{\pi}{\delta}\frac{\nu}{N},\label{eq_vertical_cut}
\end{equation}
with $\nu\in \bigl\llbracket 1, \bigl\lfloor \frac{N}{2} \bigr\rfloor \bigr\rrbracket$ and $N\geq2$. Thus, we find that the $N-1$ LSM and LBM of the $N$-layer system are obtained through vertical cuts in the bulk dispersion at quantized $q$ values given by Eq.~\eqref{eq_vertical_cut}, in the range $\left[0,\frac{\pi}{2\delta}\right]$~\footnote{Similar results could be obtain in the range $\left[-\frac{\pi}{2\delta},0\right]$ since the dispersion relation is an even function of $q$.}. 
Interestingly, Eq.~\eqref{eq_vertical_cut} suggests that the modes are confined to an effective thickness of $Nc$. Extrapolating to the single layer gives an effective thickness of $\delta$ (i.e., an interplanar distance) for one layer, as it is assumed in the literature~\cite{Yoon2009,Li2012b}, e.g. for multiple reflection calculations involving layered crystals. Let us finally note that the linear chain model outlined above can be generalized to provide a complete description of all the phonon modes in a $N$-Layer system~\cite{Luo2013,Froehlicher2015b}.

\section{Hyperspectral Raman imaging}

\begin{figure*}[!t]
\begin{center}
\includegraphics[width=0.8\linewidth]{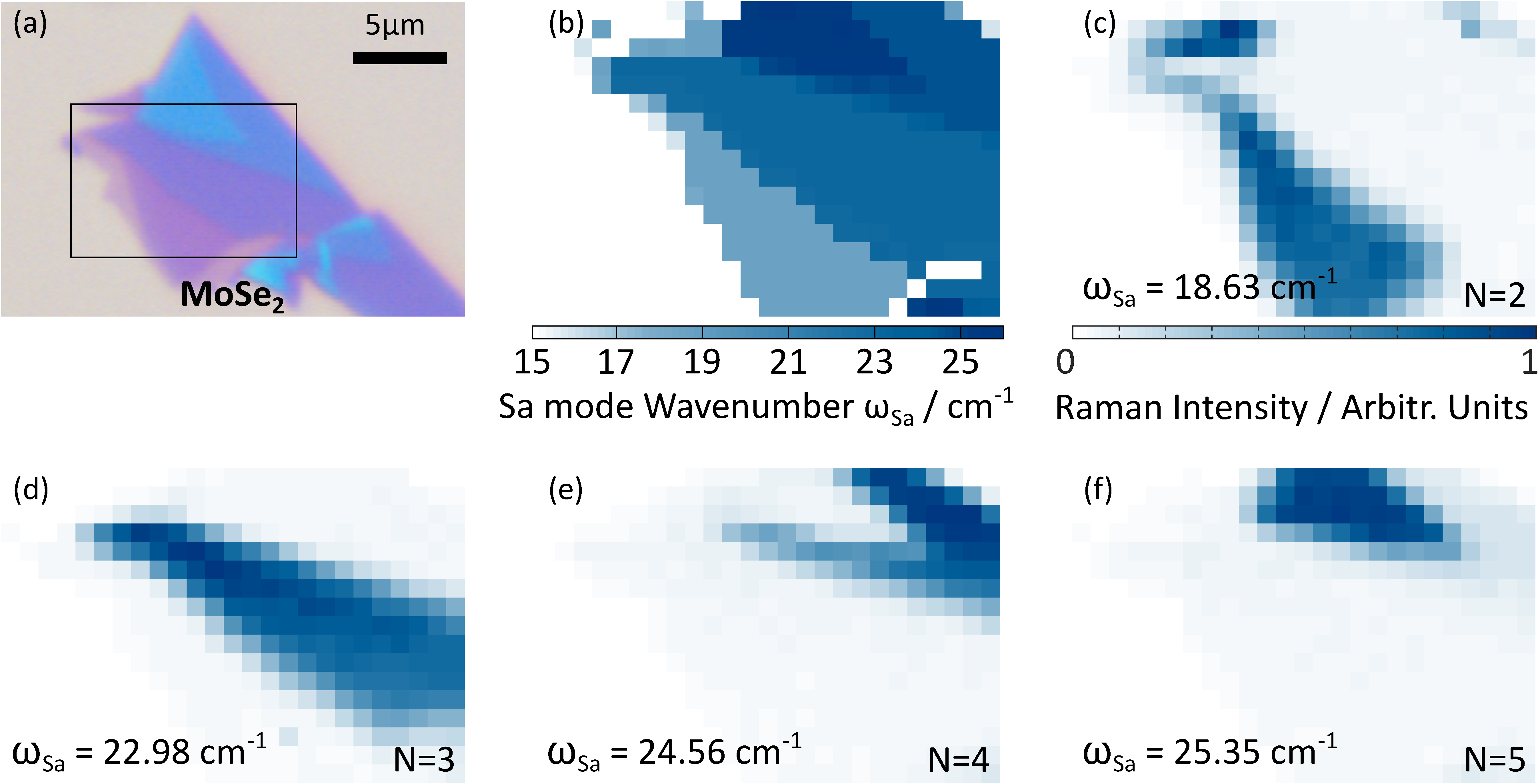}
\caption{(a) Optical image of a MoSe$_2$ flake. (b) Hyperspectral map of the wavenumber of the most prominent interlayer shear mode feature (denoted Sa in Figs.~\ref{Fig2} and \ref{Fig3}). (c-h) Hyperspectral maps of the integrated Raman intensity at selected wavenumbers that correspond to the wavenumber of the Sa feature at a given value of $N$. Measurements are performed at a laser photon energy of 2.33 eV.}
\label{Fig5}
\end{center}
\end{figure*}

\begin{figure*}[!t]
\begin{center}
\includegraphics[width=0.8\linewidth]{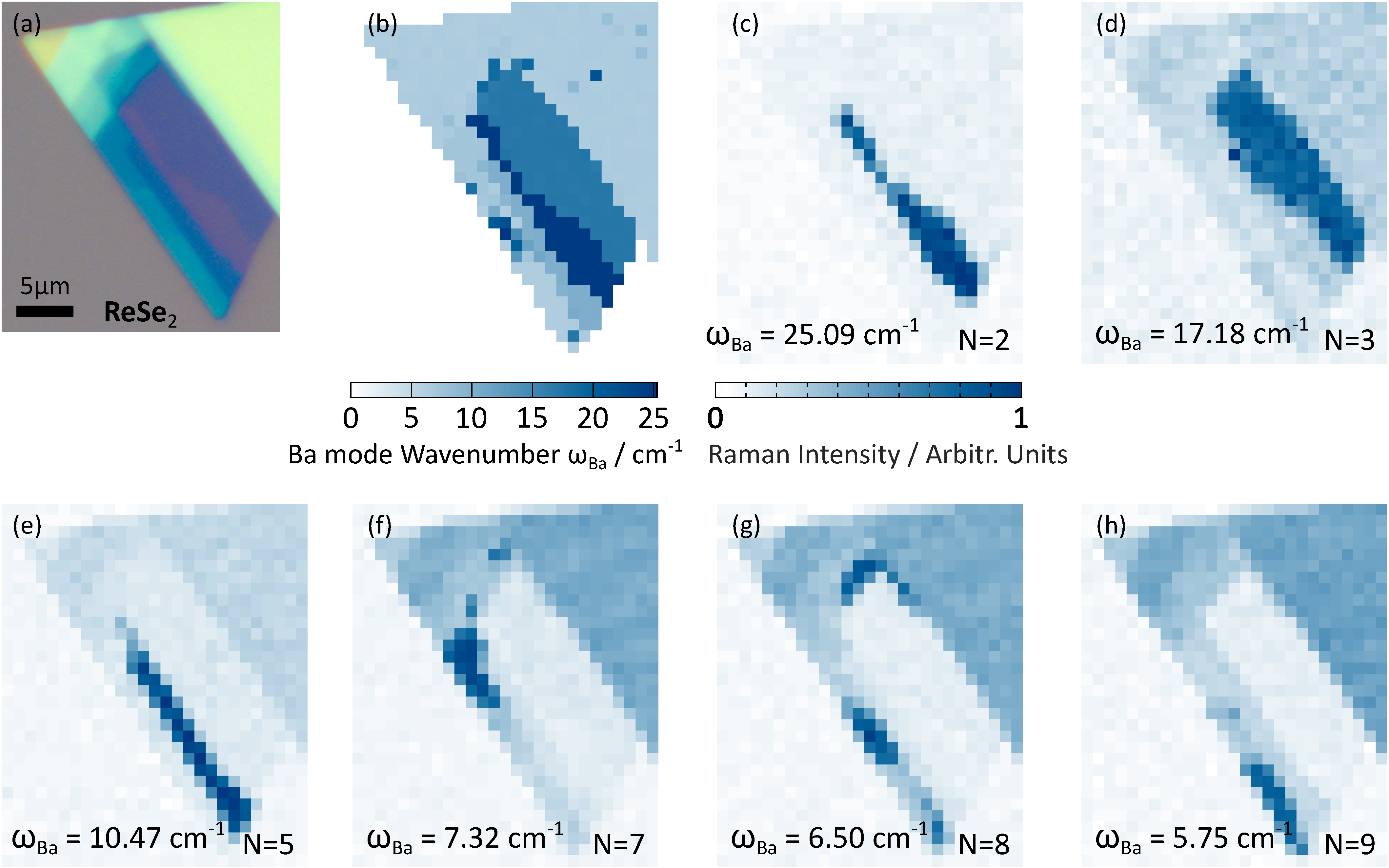}
\caption{(a) Optical image of a ReSe$_2$ flake. (b) Hyperspectral map of the wavenumber of the most prominent interlayer breathing mode feature (denoted Ba in Figs.~\ref{Fig2} and \ref{Fig3}). (c-h) Hyperspectral maps of the integrated Raman intensity at selected wavenumbers that correspond to the wavenumber of the Ba feature at a given value of $N$. Measurements are performed at a laser photon energy of 2.33 eV.}
\label{Fig6}
\end{center}
\end{figure*}

The observation of a well-defined, $N$-dependent series of LSM and LBM opens very interesting perspectives for hyperspectral Raman imaging. In Fig.~\ref{Fig4}(a), we present an optical image of a MoTe$_2$ crystal containing domains with $N$ ranging from $N=2$ to $N=13$. The number of layers is readily identified by atomic force microscopy (Fig.~\ref{Fig4}(b)) and the height difference between two MoTe$_2$ layers is measured to be $\sim 0.6~\tr{nm}$, in agreement with previous studies~\cite{Boker2001,Ruppert2014,Yamamoto2014,Lezama2015}. Note that the step between the Si/SiO$_2$ substrate and $N=2$ might be due to changes in the tip-surface interaction between the substrate and the sample~\cite{Nemes2008} or to the presence of adsorbates under the sample~\cite{Lee2010}.

In Fig.~\ref{Fig4}(c), we plot the hyperspectral map of the wavenumber $\omega_\tr{Ba}$ of the lowest energy and most intense LBM (see Fig.~\ref{Fig3}). This map readily allows one to distinguish $N$-layer domains, up to $N=13$ with a high contrast. One can also map out the Raman scattered intensity at a given shift. By selecting the Raman shifts that correspond to the Ba mode wavenumber of a $N$-layer specimen, we can then selectively image all the $N$-layer domains with an unprecedented contrast as illustrated in Fig.~\ref{Fig4}(d)-(o). Such a high contrast arises chiefly from the strong enhancement of the LBM features (especially for the Ba branch) and to the fact that the Ba branch is spectrally well separated from the other LBM and LSM modes. 

In Fig.~\ref{Fig5} and ~\ref{Fig6}, we present similar hyperspectral imaging, using the LSM in 2Hc-MoSe$_2$ layers and the LBM in 1T'-ReSe$_2$. These results establish hyperspectral Raman imaging as a swift and accurate tool for the characterization of two-dimensional materials and van der Waals heterostructures~\cite{Lui2015}.

\section{Conclusion and outlook}
We have reviewed the rigid layer shear and breathing modes in $N-$layer semiconducting transition metal dichalcogenides in the 2Hc and 1T' phases. Owing to the strong light matter-interactions in these materials, optical excitation above the optical bandgap leads to intense and rather easily measurable rigid-layer Raman features. The latter are well-separated and are very well-modeled using an elementary one-dimensional linear chain model, from which interlayer in-plane and out-of-plane force constants in the range $17-40 \times 10^{18}~\rm{N/m}^3$ and interlayer breathing force constants in the range $69-91  \times 10^{18}~\rm{N/m}^3$ are extracted. In particular, we note that for a given material, the wavenumber of the rigid layer mode show marginal sample-to-sample variation and no surface effects have to be considered to describe the rigid layer modes in substrate-supported samples. 

This article has essentially focused on the well-defined wavenumbers of the rigid layer modes. A highly stimulating direction now consists in describing the dramatic changes in the integrated intensities of the rigid layer modes. Recent Raman scattering studies using a broad range of laser photon energies have uncovered non-trivial resonant effects (see also review by Lee and Cheong in this special issue). Such resonance effects may in part be directly connected with the excitonic manifold~\cite{Chernikov2014,Wang2015,Li2015} of a given material and the symmetries of the phonon modes~\cite{Carvalho2015e,Scheuschner2015,Zhang2015b,Lee2015c,Kim2016,Soubelet2016,Song2016,delCorro2016,Tan2017}. However, quantum interference effects may largely affect the Raman intensity at a given laser photon energy in a non-intutive fashion~\cite{Miranda2017}. In the particular case of in-plane anisotropic materials, coupling between anisotropic excitons~\cite{Aslan2015,Arora2017} and complex phonon manifolds~\cite{Feng2015,Wolverson2014} result in even more complicated polarization-dependent Raman responses~\cite{Lorchat2016,Wu2015,Ribeiro2015}. All these observations show that there is still room for innovative theoretical and experimental efforts in order to provide a comprehensive description of exciton-phonon coupling in transition metal dichalcogenides and related layered materials.

\begin{acknowledgement}
We are grateful to H. Majjad for help with AFM measurements, to the StNano clean room staff for technical assistance. We acknowledge financial support from the Agence Nationale de la Recherche (under grant H2DH ANR-15-CE24-0016) and from the LabEx NIE (Under Grant ANR-11-LABX-0058-NIE). S.B. is a member of Institut Universitaire de France (IUF). 
\end{acknowledgement}


\providecommand{\latin}[1]{#1}
\providecommand*\mcitethebibliography{\thebibliography}
\csname @ifundefined\endcsname{endmcitethebibliography}
  {\let\endmcitethebibliography\endthebibliography}{}

\end{document}